# OPTICAL STOCHASTIC COONING IN TEVATRON*

V. Lebedev#, Fermilab, Batavia, IL 60510, U.S.A.

## Abstract

Intrabeam scattering is the major mechanism resulting in a growth of beam emittances and fast luminosity degradation in the Tevatron. As a result in the case of optimal collider operation only about 40% of antiprotons are used to the store end and the rest are discarded. Beam cooling is the only effective remedy to increase the particle burn rate and, consequently, the luminosity. Unfortunately neither electron nor stochastic cooling can be effective at the Tevatron energy and bunch density. Thus the optical stochastic cooling (OSC) is the only promising technology capable to cool the Tevatron beam. Possible ways of such cooling implementation in the Tevatron and advances in the OSC cooling theory are discussed in this paper. The technique looks promising and potentially can double the average Tevatron luminosity without increasing its peak value and the antiproton production.

## COOLING REQUIREMENTS

The Tevatron luminosity evolution is driven by interplay of the following major effects: the intrabeam scattering, the residual gas scattering, the RF noise and the beam-beam effects. They determine the initial luminosity lifetime of about 5-7 hours. The optimal store duration is about 16 hours and about 40% of antiprotons are burned in the particle interactions (due to luminosity). The rate of antiproton production has achieved its design value and its further growth looks extremely challenging and impossible without a major upgrade to the Antiproton source. Thus a further luminosity growth cannot be attained without beam cooling. The cooling should result in a controlled decrease of emittances so that the beams would stay at the maximum acceptable beam-beam parameter, $\xi$, in the course of entire store. That would allow us to burn in the luminosity ~80% of antiprotons and, consequently, to double the average luminosity. The required cooling times (in amplitude) are: for protons - 4 and 8 hour, and for antiprotons - 4.5 and 1.2 hour for the longitudinal and transverse degrees of freedom, correspondingly. Typical Tevatron store has $2.7 \cdot 10^{11}$ protons and $10^{11}$ antiprotons in a bunch with the rms bunch length increasing from 45 to 60 cm. Achieving the required cooling rates with stochastic cooling calls for the bandwidth of ~200 MHz which cannot be obtained in the presently tested micro-wave stochastic cooling. Electron cooling of 1 TeV (anti)protons requires ~500 MV electrons which is an expensive and extremely challenging project.

In this paper we consider a possibility of OSC suggested in Ref. [1] and later developed in Ref. [2]. Its use for the Tevatron was considered in Refs. [3] and [4].

* Work supported by the U.S. Department of Energy under contract No. DE-AC02-07CH11359
#val@fnal.gov

A suggestion to test it experimentally is reported in Ref. [5]. First we consider theory developments required for a beam optics optimization and, then, possible implementations for undulators and optical amplifiers.

Note that the OSC damps normally only horizontal and vertical degrees of freedom and the vertical cooling is achieved through the x-y coupling. In this case the horizontal motion has to be damped twice as fast resulting in the required horizontal cooling time of 4 and 0.6 hour for protons and antiprotons, correspondingly.

## TRANSFER MATRIX

The OSC of an ultra-relativistic beam assumes [1] that the beam radiates an electromagnetic radiation in a pickup undulator. Then, the radiation is amplified in an optical amplifier (OA) and produces a longitudinal beam kick in a kicker undulator. The path length difference between the light and the beam is adjusted so that a particle would interact with its own radiation. The kick is always in the longitudinal direction and the transverse cooling is achieved by coupling between transverse and longitudinal motion. The longitudinal - horizontal coupling is assumed below. The motion symplecticity binds up the transfer matrix elements so that only 10 of 16 of them are independent. In the absence of RF between points 1 and 2 the matrix between them can be expressed through the Twiss parameters of the points and the partial slip-factor between them, $\eta_{12}$, so that:

$$\mathbf{M} = \begin{bmatrix} M_{11} & M_{12} & 0 & M_{16} \\ M_{21} & M_{22} & 0 & M_{26} \\ M_{51} & M_{52} & 1 & M_{56} \\ 0 & 0 & 0 & 1 \end{bmatrix}, \quad \mathbf{x} = \begin{bmatrix} x \\ \theta_x \\ s \\ \Delta p / p \end{bmatrix}, \quad (1)$$

$$M_{11} = \sqrt{\frac{\beta_2}{\beta_1}} \left( \cos \mu + \alpha_1 \sin \mu \right),$$

$$M_{22} = \sqrt{\frac{\beta_1}{\beta_2}} \left( \cos \mu - \alpha_2 \sin \mu \right),$$

$$M_{12} = \sqrt{\beta_1 \beta_2} \sin \mu, \quad (2)$$

$$M_{21} = \frac{\alpha_1 - \alpha_2}{\sqrt{\beta_1 \beta_2}} \cos \mu - \frac{1 + \alpha_1 \alpha_2}{\sqrt{\beta_1 \beta_2}} \sin \mu,$$

$$M_{16} = D_2 - M_{11} D_1 - M_{12} D_1',$$

$$M_{26} = D_2' - M_{21} D_1 - M_{22} D_1',$$

$$M_{51} = M_{21} M_{16} - M_{11} M_{26},$$

$$M_{52} = M_{22} M_{16} - M_{12} M_{26}.$$

Here $\beta_{1,2}$ and $\alpha_{1,2}$ are the beta-functions and their negative half derivatives at the points 1 and 2, $D_{1,2}$ and $D'_{1,2}$ are the dispersions and their derivatives, and $\mu$ is the betatron phase advance between points 1 and 2. The matrix elements are enumerated similar to a 6x6 matrix but the elements related to the vertical motion (decoupled from other two degrees of freedom) are omitted. For an ultra-relativistic beam the partial slip factor is related to $M_{56}$ as:

$$\eta_{12} = \frac{M_{51}D_1 + M_{52}D_1' + M_{56}}{2\pi R} \quad (3)$$

where $R$ is the average ring radius. Note that the motion symplecticity requires $M_{56}$ sign being positive if a particle with positive $\Delta p$ moves faster than the reference particle. Substituting the matrix elements from Eq. (2) one obtains:

$$2\pi R\eta_{12} = M_{56} - D_1 D_2 \left( \frac{1+\alpha_1\alpha_2}{\sqrt{\beta_1\beta_2}}\sin\mu_1 + \frac{\alpha_2-\alpha_1}{\sqrt{\beta_1\beta_2}}\cos\mu_1 \right)$$

$$- D_1 D_2' \sqrt{\frac{\beta_2}{\beta_1}} \left( \cos\mu_1 + \alpha_1\sin\mu_1 \right)$$

$$+ D_1' D_2 \sqrt{\frac{\beta_1}{\beta_2}} \left( \cos\mu_1 - \alpha_2\sin\mu_1 \right) - D_1' D_2'\sqrt{\beta_1\beta_2}\sin\mu_1 \quad (4)$$

where $\beta_{1,2}$, $\alpha_{1,2}$, $D_{1,2}$ and $D_{1,2}'$ are the Twiss parameters at the pickup and kicker locations, correspondingly.

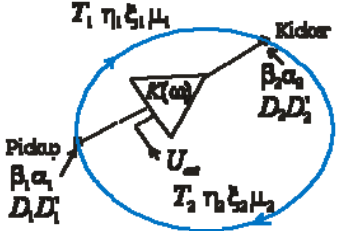

Figure 1: Layout of the cooling system

## DAMPING RATES

The layout of the cooling system is presented in Figure 1. The longitudinal kick to this particle due to interaction with its own radiation in the kicker is:

$$\frac{\delta p}{p} = \kappa \sin\left(k\,\Delta s\right) \quad (5)$$

where $k$ is the wave vector of the amplified particle wave, $\Delta s$ is the particle longitudinal displacement relative to the reference particle, and $\kappa$ is the kick maximum. Leaving only linear term in the expansion of $\sin(k\Delta s)$ in Eq. (5) and expressing $\Delta s$ through the particle positions in the pickup and the elements of the transfer matrix from pickup to kicker, $\mathbf{M}_1$, one obtains:

$$\frac{\delta p}{p} = \kappa k \left( M_{51} x_1 + M_{52}\theta_{x_1} + M_{56}\frac{\Delta p}{p} \right), \quad (6)$$

or in the matrix form

$$\boldsymbol{\delta}\mathbf{x}_2 = \mathbf{M}_c\mathbf{x}_1, \quad \mathbf{M}_c = k\kappa \begin{bmatrix} 0 & 0 & 0 & 0 \\ 0 & 0 & 0 & 0 \\ 0 & 0 & 0 & 0 \\ M_{51} & M_{52} & 0 & M_{56} \end{bmatrix} \quad (7)$$

where $\mathbf{x}_1$ is the vector of particle positions in the pickup. Taking this into account one can write down a kicker-to-kicker one turn map:

$$(\mathbf{x}_2)_{n+1} = \mathbf{M}_1\mathbf{M}_2(\mathbf{x}_2)_n + (\boldsymbol{\delta}\mathbf{x}_2)_n = (\mathbf{M}_0 + \mathbf{M}_c\mathbf{M}_2)(\mathbf{x}_2)_n, \quad (8)$$

where $n$ enumerates turns, $\mathbf{M}_2$ is the kicker-to-pickup transfer matrix, $\mathbf{M}_0 = \mathbf{M}_1\mathbf{M}_2$ is the entire ring transfer matrix, and $(\mathbf{x}_2)_n$ is related to the $n$-th turn origin beginning immediately downstream of the kicker.

The perturbation theory developed in Ref. [6] for the case of symplectic unperturbed motion yields that the tune shifts are:

$$\delta Q_k = \frac{1}{4\pi}\mathbf{v}_k^{+}\mathbf{U}\,\mathbf{M}_c\,\mathbf{U}\,\mathbf{M}_1^{T}\mathbf{U}\,\mathbf{v}_k \quad (9)$$

where $\mathbf{v}_k$ are two of four eigen-vectors of unperturbed motion chosen out of each complex conjugate pair and normalized so that $\mathbf{v}_k^{+}\mathbf{U}\,\mathbf{v}_k = -2i$ $(k=1,2)$, and

$$\mathbf{U} = \begin{bmatrix} 0 & 1 & 0 & 0 \\ -1 & 0 & 0 & 0 \\ 0 & 0 & 0 & 1 \\ 0 & 0 & -1 & 0 \end{bmatrix} \quad (10)$$

is the unit symplectic matrix. Performing matrix multiplication and taking into account that the symplecticity binds up $M_{51}$, $M_{52}$ and $M_{16}$, $M_{26}$ one finally obtains:

$$\delta Q_k = \frac{k\kappa}{4\pi}\mathbf{v}_k^{+}\begin{bmatrix} 0 & 0 & 0 & 0 \\ 0 & 0 & 0 & 0 \\ M_{126} & -M_{116} & 0 & M_{156} \\ 0 & 0 & 0 & 0 \end{bmatrix}\mathbf{v}_k \quad (11)$$

In the case of small synchrotron tune, $\nu_s \ll 1$, one can neglect the effect of RF cavities on components of the eigen-vector related to the horizontal betatron motion. Then the eigen-vector it is equal to:

$$\mathbf{v}_1 = \begin{bmatrix} \sqrt{\beta_2} \\ -(i+\alpha_2)/\sqrt{\beta_2} \\ -(iD_2(1-i\alpha_2)+D_2'\beta_2)/\sqrt{\beta_2} \\ 0 \end{bmatrix}, \quad (12)$$

Substituting Eq. (12) to Eq. (11) one obtains the damping decrement of the betatron motion:

$$\lambda_1 = -2\pi\,\mathrm{Im}\,\delta Q_1 = -\frac{k\kappa}{2}\left( D_2 M_{12,6} - D_2' M_{11,6} \right)$$

$$= -\frac{k\kappa}{2}\left( M_{156} - 2\pi R\eta_{12} \right). \quad (13)$$

The condition $\nu_s \ll 1$ also allows one to neglect the betatron motion on the synchrotron motion. Consequently, for the second eigen-vector (related to the synchrotron motion) one obtains:

$$\mathbf{v}_2 = \begin{bmatrix} -iD_2/\sqrt{\beta_s} \\ -iD_2'/\sqrt{\beta_s} \\ \sqrt{\beta_s} \\ -i/\sqrt{\beta_s} \end{bmatrix} \quad (14)$$

where $\beta_s = R\eta/\nu_s$ is the beta-function of the longitudinal motion introduced so that $\Delta s_{max} = \beta_s(\Delta p/p)_{max}$. That yields the damping decrement of the synchrotron motion:

$$\lambda_2 = -2\pi\,\mathrm{Im}\,\delta Q_2 = -\pi\,k\kappa\,R\,\eta_{12}. \quad (15)$$

Summing Eqs. (13) and (15) one obtains the sum of the decrements:

$$\lambda_1 + \lambda_1 = -\frac{k\kappa}{2}M_{156}. \quad (16)$$

## THE COOLING RANGE

The cooling force is linear for small amplitude oscillations only. Combining Eqs. (5) and (6) one obtains:

$$\frac{\delta p}{p} = \kappa \sin\left(a_x \sin(\psi_x) + a_p \sin(\psi_p)\right) \qquad (17)$$

where $a_x$, $a_p$, $\psi_x$ and $\psi_p$ are the amplitudes (expressed in the phase advance of laser wave) and phases of pickup-to-kicker path lengthening due to betatron and synchrotron motions

$$a_x \sin\psi_x = k\left(M_{1s1} x_1 + M_{1s2}\theta_{x_1}\right),$$

$$a_p \sin\psi_p = k\left(M_{1s1} D_1 + M_{1s2} D'_1 + M_{1s6}\right)\frac{\Delta p}{p}. \qquad (18)$$

Expressing $x_1$ and $\theta_{x1}$ through the particle Courant-Snyder invariant, $\varepsilon$, and introducing the amplitude of momentum oscillations, $(\Delta p/p)_{max}$, one obtains:

$$a_x = k\sqrt{\varepsilon\left(\beta_1 M_{1s1}{}^2 - 2\alpha_1 M_{1s1}M_{1s2} + \left(1+\alpha_1^2\right)M_{1s2}{}^2 / \beta_1\right)},$$

$$a_p = k\left(M_{1s1} D_1 + M_{1s2} D'_1 + M_{1s6}\right)\left(\frac{\Delta p}{p}\right)_{max}. \qquad (19)$$

Averaging momentum kicks over betatron and synchrotron oscillations one obtains the fudge factors for the transverse and longitudinal damping rates

$$\begin{bmatrix} \lambda_1(a_x, a_p)/\lambda_1 \\ \lambda_2(a_x, a_p)/\lambda_2 \end{bmatrix} \equiv \begin{bmatrix} F_1(a_x, a_p) \\ F_2(a_x, a_p) \end{bmatrix} = \begin{bmatrix} 2/(a_x \cos\psi_c) \\ 2/a_p \end{bmatrix}$$

$$\times \oint \sin\left(a_x \sin(\psi_x + \psi_c) + a_p \sin\psi_p\right)\begin{bmatrix} \sin\psi_x \\ \sin\psi_p \end{bmatrix}\frac{d\psi_x}{2\pi}\frac{d\psi_p}{2\pi}, \qquad (20)$$

where $\psi_c$ is the phase shift of the transverse cooling force. Computation of the integrals yields

$$\begin{bmatrix} F_1(a_x, a_p) \\ F_2(a_x, a_p) \end{bmatrix} = 2\begin{bmatrix} J_0(a_p)J_1(a_x)/a_x \\ J_0(a_x)J_1(a_p)/a_p \end{bmatrix}, \qquad (21)$$

where $J_0(x)$ and $J_1(x)$ are the Bessel functions. One can see that the damping rate oscillates with growth of amplitudes. For a given degree of freedom it changes the sign at its own amplitude equal to $\mu_{11}\approx3.832$ and at the amplitude of $\mu_{01}\approx2.405$ for other degree of freedom. Taking into account that the both cooling rates have to be positive for all amplitudes one obtains the stability condition, $a_{x,p} \leq \mu_{01} \approx 2.405$. That yields the stability boundaries for the emittance and the momentum spread:

$$\varepsilon_{max} = \frac{\mu_{01}{}^2}{k^2\left(\beta_p M_{1s1}{}^2 - 2\alpha_p M_{1s1}M_{1s2} + \gamma_p M_{1s2}{}^2\right)}, \qquad (22)$$

$$\left(\frac{\Delta p}{p}\right)_{max} = \frac{\mu_{01}}{2\pi k R \eta_{12}}. \qquad (23)$$

## BEAM OPTICS

To minimize the optical amplifier power $\eta_{12}$ has to be chosen as large as possible, *i.e.* at the maximum allowed by the stability boundary of Eq.(23).

$$2\pi R\eta_{12} = \frac{\mu_{01}}{k\left(\Delta p / p\right)_{max}}. \qquad (24)$$

The ratio of the cooling rates is set by the cooling scenario. That allows one to determine $M_{1s6}$. Combining Eqs. (13) and (15), and using Eq. (24) one obtains:

$$M_{1s6} = \frac{\mu_{01}\left(\lambda_1 + \lambda_2\right)}{k\lambda_1\left(\Delta p / p\right)_{max}}. \qquad (25)$$

The Tevatron cooling scenario implies the initial rms momentum spread and the rms normalized emittance being equal to $1.2\cdot10^{-4}$ and $3.3$ mm mrad, correspondingly. Requiring $4\sigma$ and $5\sigma$ cooling ranges for the longitudinal and transverse motions, correspondingly, one obtains: $\gamma\varepsilon_{max} = 83$ mm mrad and $(\Delta p/p)_{max} = 4.8\cdot10^{-4}$. Corresponding values of $M_{1s6}$ and $\eta_{12}$ for equal damping rates ($\lambda_1 = \lambda_2$) are shown in Table 1 for the wavelengths of 2 and 12 μm[*].

Table 1: Major optics parameters

| Optical amplifier wavelength [μm] | 2 | 12 |
|---|---|---|
| $M_{56}$ [mm] | 3.2 | 19.2 |
| $2\pi R \eta_{12}$ [mm] | 1.6 | 9.6 |
| Total chicane length [m] | 69.6 | 59.3 |
| $\Delta D$ for 10% damping rate change [cm] | 0.45 | 1.7 |
| $\Delta D'$ for 10% damping rate change[$10^{-3}$] | 2 | 1.2 |

A particle and its radiation from the pickup undulator have to arrive to the kicker undulator simultaneously. But an optical amplification results in a delay of the beam signal. The study presented in Ref. [2] shows that a delay of 5 to 10 mm is required. To compensate this delay the beam path lengthening by a four-dipole chicane was proposed in Ref. [1]. The 5.3 mm delay created by a chicane with 6 T dipoles is implied in the below estimates. In the absence of quadrupole focusing in the chicane its delay, $\Delta L$, and $M_{56}$ are approximately related so that: $M_{56} \approx 2\Delta L$. It also results in that $M_{56}$ and $2\pi R\eta_{12}$ are equal and, consequently, there is no horizontal damping. Therefore focusing in the chicane is required. As one can see from Table 1 the required $M_{56}$ is almost 4 times larger than $\Delta L$ for 12 μm wavelength and moderate focusing is sufficient. For the case of 2 μm wavelength the required $M_{56}$ is about 1.5 times smaller than $\Delta L$ and strong focusing is required. Figure 2 presents the beta-functions and dispersions in the chicane for both cases. For both of them the dispersion is much smaller than the dispersion in the Tevatron utility straights and additional quads will be required to match the cooling section to the Tevatron optics. In this example a periodic solution for beta-functions was used. For a practical proposal an additional beta-function adjustment is required so that the sample lengthening described by the top Eq. (19) would be optimized. The lengthening should not be too small, which results in an excessive optics sensitivity, and should not be too large so that particles with large betatron amplitudes would be cooled (see Eq. (22)).

$M_{56}$ for the chicane depends only on its structure. It is not affected by the rest of the ring optics and therefore is

---

[*] Note that the damping in both planes requires the signs of $M_{1s6}$ and $\eta_{12}$ to be the same. They can be changed by changing the cooling system phase by 180 deg.

quite stable. Consequently, the sum of cooling rates ($\lambda_1 + \lambda_2$) is stable too. In contrary, the chicane partial slip factor, $\eta_{12}$, is determined by the dispersion in the chicane and is strongly affected by the ring optics. A requirement to keep the damping rates within 10% results in a dispersion accuracy at the chicane entrance being 1.7 and 0.45 cm for 12 and 2 µm wavelengths, correspondingly (see Table 1.) It is about an order of magnitude better than the present optics accuracy. Its further improvement, in particular for 2 µm option, presents a very challenging task.

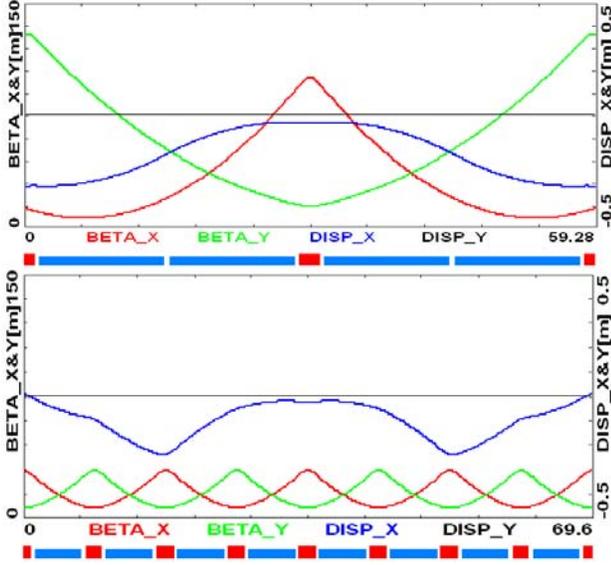

Figure 2: Beta-functions and dispersions in the cooling chicane for optics optimized for 12 (top) and 2 (bottom) µm optical amplifiers.

## KICKER UNDULATOR

Because of large relativistic factor of the Tevatron beam ($\gamma = 1045$) the period of undulator is large: ~1.5 m for 2 µm and ~7 m for 12 µm wavelength. Therefore a large number of wiggles cannot be used. Four types of magnetic field configurations were considered: (1) kick in a SC Tevatron dipole, (2) a wiggler build with alternating sign dipoles, (3) a standard harmonic undulator, and (4) a helical dipole.

The electric field of electromagnetic wave propagating along $z$ axis and polarized in the $x$-plane can be expressed in the following form

$$E_x(x, y, z, t) = \mathrm{Re}\left( E_0 e^{i(\omega t - kz)} \frac{\sigma_\perp^2}{\sigma^2(z)} \exp\left(-\frac{1}{4} \frac{x^2 + y^2}{\sigma^2(z)}\right) \right),$$

$$E_y(x, y, z, t) = 0,$$

$$E_z(x, y, z, t) = \mathrm{Re}\left( iE_0 e^{i(\omega t - kz)} \frac{\sigma_\perp^2 x}{4k\sigma^4(z)} \exp\left(-\frac{1}{4} \frac{x^2 + y^2}{\sigma^2(z)}\right) \right),$$

(26)

where $E_0 = \sqrt{4P/c\sigma_\perp^2}$ is the electric field in the waist, $\sigma^2(z) = \sigma_\perp^2 - iz/2k$, $k = 2\pi/\lambda_w$, $\lambda_w$ is the wavelength, and $\sigma_\perp$ is the rms size at the waist (power density). The longitudinal kick was obtained by numerical integration of the following equation, $\Delta E = \int(\mathbf{E} \cdot \mathbf{v}) dt$, along the particle trajectory in the magnetic field. For all cases $\sigma_\perp$ and the wave waist offset in the $x$ and $y$ planes were adjusted to maximize the kick value. Note that in the case of short undulator or dipole both $E_x$ and $E_z$ components make comparable contributions to the integral and have to be accounted.

Making a harmonic magnetic field ($B_y \propto \sin(k_{wgl}z)$) with period of many meters is not a practical engineering choice. Therefore a wiggler consisting of dipoles with constant magnetic field but changing polarity is considered. To separate the light and particle beams there are also dipoles immediately adjacent to the wiggler. They have the same polarity and strength as the outer wiggler dipoles. Note that the harmonic undulator has the same kicker efficiency per unit length as the dipole wiggler and therefore its use does bring any advantages.

A kick in a dipole (implying no wiggler at all) does not require additional space and therefore can be an attractive option. Although there are no wiggles in a dipole a tight focusing of the wave and its offset from the beam center in the horizontal plane allow one to obtain a considerable kick. As can be seen in Figure 3 there is little gain if a wiggler consisting of three dipoles (total length of ~25 m for 4 T dipoles and $\lambda_w$ = 12 µm) is used. A wiggler consisting of five dipoles is about 2 times more efficient but requires ~40 m space per wiggler for $\lambda_w$ = 12 µm. Taking into account ~60 m required for the chicane it is a maximum space which can be allocated for the wiggler.

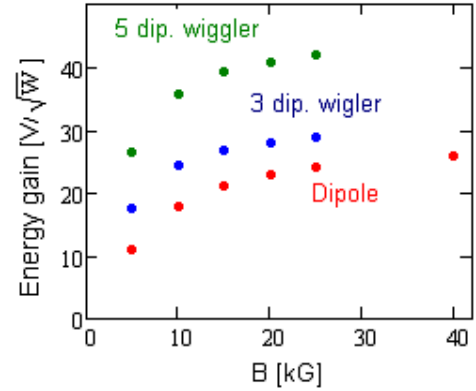

Figure 3: Dependence of kicker efficiency on the magnetic field for 12 µm wavelength with a kick in a dipole (red) and in wigglers build from 3 (blue) and 5 (red) dipoles.

A helical undulator with a circular polarized electromagnetic wave is ~$\sqrt{2}$ times more efficient than the flat ones. For large number of periods its kick amplitude can be expressed by the following expression

$$\frac{\Delta E_{max}}{e} \approx \sqrt{8.837 n_{wgl} P Z_0} \frac{K_u^2}{1 + K_u^2},$$

(27)

where $e$ is the electron charge, $n_{wgl}$ is the number of wiggler periods, $Z_0$ is the free space impedance, $K_u = eB\lambda_{wgl}/(2\pi mc^2)$ is the undulator parameter, and $\lambda_{wgl}$ is its period. The above equation implies that the radiation is focused into the rms spot size equal to $\sigma_\perp \approx \sqrt{0.473 L \lambda_w}$, and the wavelength and the period of undulator are related so that $\lambda_{wgl} = 2\gamma^2 \lambda_w / (1 + K_a^2)$. As can be seen in Figure 4 the efficiency is somewhat smaller than predictions of Eq. (27) for small number of periods.

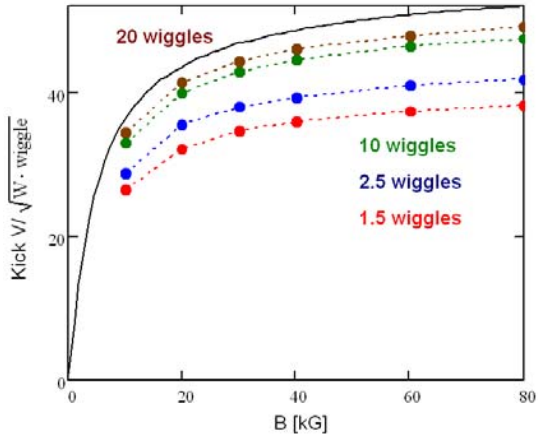

Figure 4: Dependence of helical undulator efficiency on the magnetic field and the number of periods for $\lambda_w = 12$ μm. The black line represents an asymptotic for $n_{wgl} \rightarrow \infty$.

Table 2: Parameters of possible cooling schemes

| $\lambda_w$ [μm] | Wiggler type/$n_{wgl}$ | B [T] | Length [m] | $G_{kick}$ [eV/$\sqrt{W}$] | $P_l$† [W] |
|---|---|---|---|---|---|
| 12 | Tevatron dipole /(N/A) | 4.2 | N/A | 26 | 125 |
| 6 | | | | 18 | 133 |
| 2 | | | | 14 | 71 |
| 12 | HD/2.5 | 2 | 40 | 56 | 28 |
| | HD/8 | 8 | 44 | 132 | 5 |
| 6 | HD/7 | 6 | 38 | 110 | 3.5 |
| 2 | HD/12 | 6 | 36 | 116 | 1.05 |

* HD – helical dipole

† It is the average power. The peak power is about 100 times higher

## POWER OF OPTICAL AMPLIFIER

Expressing $\kappa$ in Eq. (15) through the kick amplitude one can express the damping rate in the following form

$$\lambda_2 = \frac{1}{2} \frac{\Delta E_{max}}{E_0 \sigma_p} \frac{\mu_{01}}{n_{\sigma p}} F_2 \left( \frac{\mu_{01}}{n_{\sigma x}}, \frac{\mu_{01}}{n_{\sigma p}} \right) \quad (28)$$

where $n_{\sigma p}$ and $n_{\sigma x}$ are the cooling ranges (expressed in the beam σ's) for longitudinal and transverse degrees of freedom, $E_0$ is the beam energy, and $\sigma_p$ is the relative rms momentum spread.

Assuming that the dependence of optical amplifier gain on the frequency can be described by a Gaussian we obtain the average power of laser amplifier:

$$P_l = \frac{n_b N_p f_0}{\Delta f_{FWHM}} \sqrt{\frac{\ln(2)}{\pi}} \left( \frac{\Delta E_{max}}{e G_{kick}} \right)^2 \quad (29)$$

where $n_b$ is the number of bunches, $N_p$ is the number of particles per bunch, $f_0$ is the revolution frequency, $\Delta f_{FWHM}$ is the bandwidth of OA (FWHM), and $G_{kick}$ is the kicker efficiency introduced so that $\Delta E_{max}/e = G_{kick}\sqrt{P}$. The results of calculations are summarized in Table 2, where we assume that $n_b = 36$, $N_p = 3 \cdot 10^{11}$, $\sigma_p = 1.2 \cdot 10^{-4}$, the relative optical amplifier bandwidth of 6% (FWHM), and the longitudinal damping time, $(\lambda_2 f_0)^{-1}$, equal to 4.5 hour corresponding to the rms single particle kick $\Delta E_{max} = 0.66$ eV. As one can see a usage of Tevatron dipole as a kicker does not look attractive because of too high power of laser amplifier. However a dipole can be used instead of pickup undulator in the case of insufficient space. It requires additional ~15 Db gain for the OA.

## DISCUSSION

Potentially, the OSC allows one to double the average Tevatron luminosity. The system can be located in the Tevatron C0 straight section which has sufficient space. Its installation requires significant investment and downtime. In particular it requires a modification of beam optics which includes: new quadrupoles and new quad circuits for existing ones, a relocation of existing and installation of new dipoles. The optics work will be complicated by a requirement to keep the same fractional tunes and to support helical orbits separating protons and antiprotons in the cooling section. Cooling of protons was only discussed above but doubling the luminosity integral also implies aggressive cooling of antiprotons. Their cooling time should be ~4 times faster; but because they normally have 4 times smaller intensity the same power of OA is required.

The 2 μm wavelength looks attractive because the OA was already demonstrated [7] but that requires very high accuracy of optics control. A 12 μm OA considered in Ref. [2] with its first tests reported in Ref. [8] requires an additional investment and considerable time for further development. For both wavelengths the power of OA stays below 10 W if the helical undulator is used.

This study showed that there is no fast way (2-3 years) to introduce the OSC in Tevatron so that it could be implemented in the course of the Tevatron Run II. However it demonstrates a high potential of the OSC for the Tevatron luminosity increase.

*Acknowledgements* The author is grateful to M. Zolotarev and A. Zholents for constructive and extremely useful discussions.